\documentstyle[twoside,fleqn,espcrc2,fps]{article}

\newcommand{\AmS}{{\protect\the\textfont2
  A\kern-.1667em\lower.5ex\hbox{M}\kern-.125emS}}

\title{Chiral transition in a strongly coupled fermion-gauge-scalar model}

\author{Xiang-Qian Luo\address{HLRZ, Forschungszentrum, 
D-52425 J\"ulich, Germany\\
and Deutsches Elektronen-Synchrotron DESY, D-22603 Hamburg, Germany 
}
\thanks{Speaker.
}
and Wolfgang Franzki\address{Institute of Theoretical Physics E, 
RWTH Aachen, D-52056
  Aachen, Germany
}
}

\begin{document}

\begin{abstract}
We report the recent results from
the computer simulations of a fermion-gauge-scalar model
with  
dynamical chiral-symmetry breaking 
and chiral transition induced by the scalar field.
This model might be considered 
to be a possible alternative to the Higgs mechanism
of mass generation.
A new scheme is developed for detecting the chiral transition. 
Our results show with higher precision than the earlier works
that
the chiral transition line joins the Higgs phase transition line, 
separating the Higgs and Nambu
(chiral-symmetry breaking)
phases. 
The end point of the Higgs transition
with divergent correlation lengths is therefore
suitable for an investigation of the continuum limit.
\end{abstract}

\maketitle

\section{INTRODUCTION}

Some strongly coupled lattice fermion-gauge models
with a charged
scalar field, which break chiral symmetry dynamically,
might be considered to be a possible alternative to the Higgs mechanism
for mass generation, as discussed in \cite{Jersak,FrJer}.

Let us concentrate on a prototype
with $U(1)$ gauge group, a 
scalar of fixed modulus and one staggered fermion (corresponding to 4 flavors),
where both the scalar and fermion have charge one.
The action has been described in \cite{Jersak,FrJer} with three
bare parameters $(\beta,\kappa,m_0)$. The dynamical mass generation
is meaningful only in the chiral limit $m_0=0$. 
We consider here the phase transition line {\bf NET} between two 
phases \cite{Jersak,FrJer}:

\noindent
(1) Dynamical mass generation 
(Nambu) phase, below the {\bf NET} line,
where chiral symmetry is spontaneously broken
($\langle\bar{\psi} \psi\rangle \not=0$) due to
the strong gauge fluctuations so that the
fermion mass $m_F$ is dynamically generated;

\noindent
(2) Higgs phase, above the {\bf NETS} line,
where the Higgs mechanism is operative, but
$\langle\bar{\psi} \psi\rangle=m_F=0$.

\noindent
The scalar field induces a second order chiral phase transition {\bf NE}
line which opens the possibility for approaching the 
continuum.

Whether such a model can replace the Higgs mechanism depends crucially on
the existence and renormalizability 
of the continuum limit.  To search for
such a continuum theory and grasp its nature, we need
to make precise determination of the second order phase transition point
with divergent correlation lengths.
For such a purpose,
we have done extensive simulations 
using Hybrid Monte Carlo
(HMC) algorithm and developed some new methods 
for locating the 
{\bf NE} line.

\section{HMC SIMULATIONS}
The HMC simulations have been done
on $6^3 16$ and $8^3 24$, where on $6^3 16$,
we have better statistics (1024-6500 trajectories)
for different $(\beta,\kappa,m_0)$.
The detailed results for the spectrum are reported in \cite{FrLuo}.
We have measured the following local observables: plaquette energy $E_p$,
link energy $E_l$ and chiral condensate $\langle{\bar \psi} \psi\rangle$, 
where for $\langle{\bar \psi} \psi\rangle$ we use the stochastic estimator method.
However, 
it is very difficult
to use the local quantities at finite $m_0$ to detect 
a critical behavior on the {\bf NE} line, since they
show smooth behavior as a function of $\beta$ or $\kappa$.
(One could expect
the critical behavior only in the infinite volume and chiral limit.)
For $(\beta,\kappa)$ near the point {\bf E}, the peaks of
susceptibility
for different quantities develop and coincide, while
the boson mass $am_S$ gets smaller.
Concerning the location of the {\bf ET} line,
on the $6^3 16$ and $8^3 24$ 
for $\kappa < 0.31$ or $\beta>0.64$ and $m_0=0.04$,
we find explicit two state signals from the thermo-cycle, 
time history and histogram
analysis of the local quantities. 

\begin{figure}[htb]
\fpsxsize=7.5cm
\fpsbox[70 90 579 760]{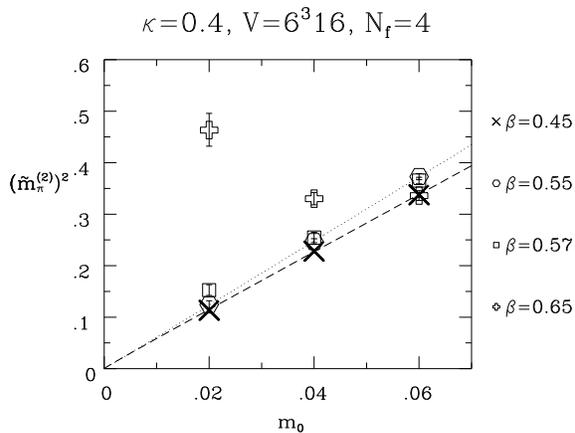}
\vspace{-5mm}
\caption{The pseudoscalar (pion) mass squared versus bare fermion
  mass.
For large $\beta$, large contributions from the
excited states are found and systematic errors are expected.
}
\label{fig1}
\end{figure}

On the {\bf NE} line, the
$\pi$ meson shows more obviously
the phase transition than other quantities.
In the Nambu phase, the $\pi$ meson should obey the PCAC relation.
In the symmetric phase, the $\pi$ meson is no longer a Goldstone
boson, and one should observe a deviation from PCAC.
At $\kappa=0.4$, these properties 
are nicely seen in fig.~\ref{fig1},
from which one sees that 
for $\beta < 0.57$ where the system is in the
broken phase, we have Goldstone bosons.
However,
on $6^3 16$, even 
at $\beta=0.57$ (possibly in the chiral symmetric phase), 
a linear extrapolation leads to
$\langle{\bar \psi} \psi\rangle\vert_{m_0=0} \approx 0.13$. 
For larger $\beta$, the extrapolated result gets smaller (e.g.
at $\beta=0.65$, $\langle{\bar \psi} \psi\rangle\vert_{m_0=0} \approx 0.05$)
and is expected to vanish in the $V \to \infty$
limit. Of course, one should not expect the linear extrapolation to be
valid at the critical point.

\section{NEW ORDER PARAMETER}

$\langle\bar{\psi} \psi\rangle$ is not a convenient 
order parameter for the chiral transition of a finite
system due to the sensitivity of chiral extrapolation. 
We employ a
different method for determining the chiral transition, namely we
calculate the chiral susceptibility in the chiral limit,
defined by
\begin{eqnarray}
\chi_{chiral}= 
{\partial  \langle\bar{\psi} \psi\rangle \over \partial m_0} \vert_{m_0=0}.
\label{def}
\end{eqnarray}
If there is a  second order chiral phase transition, 
$\chi_{chiral}$
should
be divergent (in other words, 
$\chi_{chiral}^{-1}$ should be zero)
at the critical point and in the thermodynamical limit.
In the chiral limit, the chiral susceptibility 
in the Nambu phase is difficult to obtain, but it 
is calculable in the chiral symmetric phase \cite{Azcoiti}.
It can be shown that in the symmetric phase $\chi_{chiral}^{-1}$,
defined in eq. (\ref{def}), is the same as
\begin{eqnarray}
\chi^{-1}=\left \{ {2 \over V}
\langle  \sum_{i=1}^{V/2} {1 \over \lambda_{i}^2}\rangle\vert_{m_0=0} \right \}^{-1},
\label{order}
\end{eqnarray}
where $\lambda_{i}$ are the positive eigenvalues of the massless
fermionic matrix.
Approaching the {\bf NE} line from the symmetric phase
by fixing $\kappa$, 
$\chi^{-1}$ should behave as
$
\chi^{-1} \propto (\beta-\beta_c)^{\gamma},
$
corresponding to the divergent correlation length at the
second order phase transition point
in the thermodynamical limit $V \to \infty$. 
In the Nambu phase, it can also be shown that 
eq. (\ref{order}) is equivalent to
\begin{eqnarray}
\chi^{-1} 
=\left\{     {V \over 2}(\langle\bar{\psi} \psi\rangle\vert_{m_0=0})^2 \right\}^{-1}
\end{eqnarray}
in the $V \to \infty$ limit. Then in such a limit,
$\chi^{-1}$ should be zero since $\langle\bar{\psi} \psi\rangle\vert_{m_0=0} \not=0$
in the Nambu phase.

Therefore, $\chi^{-1}$ defined in eq. (\ref{order}) is a suitable order parameter
for the chiral phase transition:
it is zero in the broken phase, and it is nonzero in the symmetric phase.

\begin{figure}[htb]
\fpsxsize=7.5cm
\fpsbox[70 90 579 760]{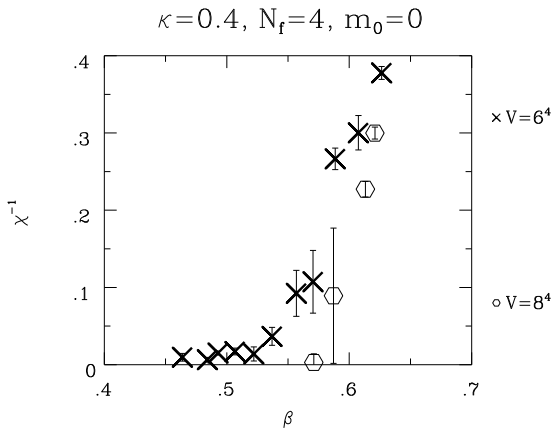}
\vspace{-5mm}
\caption{The order parameter for the chiral transition.}
\label{fig2}
\end{figure}

Let us again focus on the results 
at $\kappa=0.4$.
To perform
the calculation,
we generalize MFA \cite{MFA},
in which the chiral limit $m_0=0$
is accessible,
to the fermion-gauge-scalar models.
From
fig.~\ref{fig2}, we observe that on $8^4$ the chiral transition appears at
$\beta_c \approx 0.57$,
being consistent with the observation of fig.~\ref{fig1}.

\begin{figure}[htb]
\fpsxsize=7.5cm
\fpsbox[70 90 579 760]{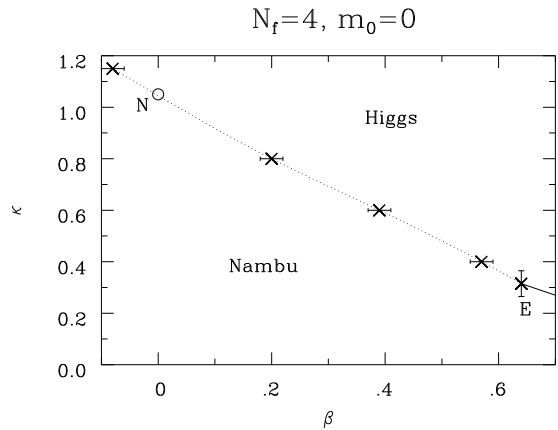}
\vspace{-5mm}
\caption{Location of the {\bf NE} line on finite lattices.}
\label{fig3}
\end{figure}

\section{DISCUSSIONS} 
The location of the  {\bf NE} line 
on the available lattices
obtained by the above methods
is summarized in fig.~\ref{fig3}, 
where the point {\bf N} is plotted by interpolation.
We have determined the phase transition line
{\bf NE} with high precision and demonstrated that this
second order
chiral transition line joins the Higgs phase transition line at the
end point {\bf E} being around $(\beta,\kappa) = (0.64, 0.31)$, 
separating the Higgs and Nambu phases.
No finite size scaling analysis has been done, and
larger lattices are required for such a purpose.

From the spectroscopy \cite{FrLuo}, we know that
$am_F$ scales to zero when crossing
the chiral transition line {\bf NE}. Nevertheless,
the susceptibility for $E_l$ and correlation length
for the composite scalar $(am_S>1.5)$ 
remain finite on the whole 
{\bf NE} line except approaching the end point {\bf E}.
Therefore,  the end point, hopefully being a second
order point with divergent correlation lengths, is the
most suitable candidate for the continuum limit.

Further work to be done is to study the finite size effects,
analyze the dependence of the end point {\bf E} on the bare
fermion mass, 
investigate the scaling properties
and understand the nature of the end point, which is underway \cite{FFJL}.

We would like to thank C. Frick and J. Jers\'ak for collaboration
and V. Azcoiti for useful discussions.
The HMC simulations have been performed on HLRZ Cray Y-MP8/864
and NRW state SNI/Fujitsu VPP 500.

\end{document}